\documentclass[12pt]{article}
\usepackage{epsfig}

\newcommand\pubnumber{}
\newcommand\pubdate{}

\def\support{\footnote{On behalf of the BaBar Collaboration}}

\def\padova{Universit\`a di Padova and INFN sezione di Padova\\
Padova, ITALY}

\def\Title#1{\begin{center} {\Large #1 } \end{center}}
\def\Author#1{\begin{center}{ \sc #1} \end{center}}
\def\Address#1{\begin{center}{ \it #1} \end{center}}

\newcommand\pubblock{\rightline{\begin{tabular}{l} \pubnumber\\
         \pubdate  \end{tabular}}}
\newenvironment{Abstract}{\begin{quotation}  }{\end{quotation}}
\newenvironment{Presented}{\begin{quotation} \begin{center} 
             PRESENTED AT\end{center}\bigskip 
      \begin{center}\begin{large}}{\end{large}\end{center} \end{quotation}}

\def\beq{\begin{equation}}
\def\eeq#1{\label{#1}\end{equation}}
\def\eeqn{\end{equation}}

\def\beqa{\begin{eqnarray}}
\def\eeqa#1{\label{#1}\end{eqnarray}}
\def\eeqan{\end{eqnarray}}

\let\bar=\overbar

\def\Dslash{\not{\hbox{\kern-4pt $D$}}}
\def\dslash{\not{\hbox{\kern-2pt $\del$}}}

\def\BR{\mbox{\rm BR}}

\def\msb{{\bar{\ssstyle M \kern -1pt S}}}

\input babarsym
\newcommand{\bi}{\begin{itemize}}
\newcommand{\ei}{\end{itemize}}
\newcommand{\ben}{\begin{enumerate}}
\newcommand{\een}{\end{enumerate}}
\newcommand{\bc}{\begin{center}}
\newcommand{\ec}{\end{center}}
\newcommand{\bt}{\begin{table}}
\newcommand{\et}{\end{table}}
\newcommand{\be}{\begin{equation}}
\newcommand{\ba}{\begin{eqnarray}}
\newcommand{\ea}{\end{eqnarray}}

\newcommand{\la}{\ifmmode {\leftarrow} \else {$\leftarrow$}\fi}
\newcommand{\Ra}{\ifmmode {\Rightarrow} \else {$\Rightarrow$}\fi}
\newcommand{\La}{\ifmmode {\Leftarrow} \else {$\Leftarrow$}\fi}
\newcommand{\Lra}{\ifmmode {\Longrightarrow} \else {$\Longrightarrow$}\fi}
\newcommand{\Lla}{\ifmmode {\Longleftarrow} \else {$\Longleftarrow$\fi}}
\newcommand{\Llra}{\ifmmode {\Longleftrightarrow} \else {$\Longleftrightarrow$\fi}}
\newcommand{\Lk}{\ifmmode {{\cal L}} \else {${\cal L}$}\fi}
\newcommand{\Wt}{\ifmmode {{\cal W}} \else {${\cal W}$}\fi}
\newcommand{\Br}{\ifmmode {{\cal B}} \else {${\cal B}$}\fi}
\newcommand{\N}{\ifmmode {{\cal N}} \else {${\cal N}$}\fi}
\newcommand{\G}{\ifmmode {{\cal G}} \else {${\cal G}$}\fi}
\newcommand{\E}{\ifmmode {{\cal E}} \else {${\cal E}$}\fi}
\newcommand{\Pfr}{\ifmmode {{\cal F}} \else {${\cal F}$}\fi}
\newcommand{\Aone}{\ifmmode {{\cal A}_1} \else {${\cal A}_1$}\fi}
\newcommand{\rha}{\ifmmode{\mbox{\rho^2_{{\cal A}_1}}} \else {\mbox{$\rho^2_{{\cal A}_1}$}}\fi}
\newcommand{\rhf}{\ifmmode{\rho^2_{\cal F}}\else{\mbox{$\rho^2_{\cal F}$}}\fi}
\newcommand{\om}{\ifmmode {w} \else {$w$}\fi}
\newcommand{\dom}{\ifmmode {\Delta w} \else {$\Delta w$}\fi}

\newcommand{\tBz}{\ifmmode {\tau_{\Bz}} \else {$\tau_{\Bz}$}\fi}
\newcommand{\tBp}{\ifmmode {\tau_{\Bu}} \else {$\tau_{\Bu}$}\fi}

\newcommand{\psoft}{\ifmmode {\pi_{s}} \else {$\pi_{s}$}\fi}
\newcommand{\plab}{\ifmmode{p} \else {$p$} \fi}
\newcommand{\ctdl}{\ifmmode{ \cos(\theta_{\Dstar\ell}) } \else {$\cos(\theta_{\Dstar\ell})$} \fi}
\newcommand{\ks}{\ifmmode{k^*} \else {$k^*$} \fi}
\newcommand{\mnutag}{\ifmmode{m^2_{\nu ,tag}} \else {$m^2_{\nu ,tag}$}\fi} 
\newcommand{\mnusig}{\ifmmode{m^2_{\nu ,sig}} \else {$m^2_{\nu ,sig}$}\fi} 
\newcommand{\DTau}{\ifmmode {\Delta \tau} \else {$\Delta \tau$}\fi}
\newcommand{\ggcc}{\ifmmode {GeV^2/c^4} \else {$GeV^2/c^4$}\fi}
\def\BpBm {\ensuremath{B^+ {\kern -0.16em \Bub}}}

\def\poverq2{\ensuremath{\bigg\vert\frac{p}{q}\bigg\vert^2}\xspace}
\def\qoverp2{\ensuremath{\bigg\vert\frac{q}{p}\bigg\vert^2}\xspace}

\def\BzBz     {\ensuremath{\mbox{\Bz {\kern -0.1em \Bz}}}\xspace}
\def\BzBzb     {\ensuremath{\mbox{\Bz {\kern -0.1em \Bzb}}}\xspace}
\def\BzbBzb   {\ensuremath{\mbox{\Bzb {\kern -0.1em \Bzb}}}\xspace}

\def\Kp{\ensuremath {K^{+}\xspace}}
\def\Km{\ensuremath {K^{-}\xspace}}

\def\qoverp{\ensuremath{\frac{q}{p}}}

\begin{document}
\begin{titlepage}
\pubblock

\vfill
\Title{Semileptonic Mixing Asymmetry Measurements of $A^d_{SL}$ and $A^s_{SL}$}
\vfill
\Author{ Martino Margoni\support}
\Address{\padova}
\vfill
\begin{Abstract}
Standard Model predictions of the CP violation in the 
mixing of $B^0_d$ and $B^0_s$ mesons are beyond the present experimental sensitivity, 
any observation would be therefore a hint of new physics. 

The D0 collaboration measures a value of the semileptonic mixing asymmetry for a 
mixture of $B^0_d$ and $B^0_s$ mesons, $A^b_{SL}$, which misses the Standard Model expectation by 3.9 
standard deviations.
The world averages of the flavor specific measurements of the semileptonic asymmetries for
$B^0_d $ and $B^0_s$ mesons, $A^d_{SL}$ and $A^s_{SL}$, are instead in agreement with the Standard Model. 

The combination of the various $A^q_{SL}~(q=d, s)$ measurements and the recent LHCb results 
on $B^0_s\rightarrow J/\psi\phi$ have placed
tight bounds on the hypothesis of new physics which can explain the D0 result.
\end{Abstract}
\vfill
\begin{Presented}
the 2013 Flavor Physics and CP Violation (FPCP-2013)\\
Buzios, Rio de Janeiro, Brazil, May 19-24 2013
\end{Presented}
\vfill
\end{titlepage}
\def\thefootnote{\fnsymbol{footnote}}
\setcounter{footnote}{0}

\section{Introduction}
\label{sec:intro}
The neutral $B^0_q~(q=d,s)$ mesons mix with their antiparticles leading to
oscillations between the mass eigenstates. The time evolution of the neutral
mesons doublet is governed by an effective hamiltonian $H=M-\frac{i}{2}\Gamma$, where $M$ is the mass matrix and $\Gamma$ 
is the decay matrix, from which the light (L) and heavy (H) physical eigenstates with defined masses and widths 
are obtained:
\begin{eqnarray}
|B^{L,H}_q>=\frac{1}{\sqrt{1+|q/p|_q^2}}(|B_q>\pm(q/p)_q|\bar{B_q}>).
\nonumber
\end{eqnarray}
The two eigenstates have a mass difference
$\Delta m_q= m^H_q-m^L_q \simeq |M^q_{12}|$ and a total decay width difference $\Delta \Gamma=\Gamma^L_q-\Gamma^H_q\simeq |\Gamma^q_{12}|\cos \phi_q$,
where $\phi_q=\arg(-M^q_{12}/\Gamma^q_{12})$ is a CP violating phase.

The time-independent CP violation asymmetry in the $B^0_q$ mixing is defined as
\begin{eqnarray}
\nonumber
A^q_{CP}=\frac{{\rm Prob}(\bar{B^0_q}(0)\rightarrow B^0_q(t))-{\rm Prob}(B^0_q(0)\rightarrow \bar{B^0_q}(t))}{{\rm Prob}(\bar{B^0_q}(0)\rightarrow B^0_q(t))+{\rm Prob}(B^0_q(0)\rightarrow \bar{B^0_q}(t))}=\frac{1-|q/p|_q^4}{1+|q/p|_q^4}=
\frac{|\Gamma^q_{12}|}{|M^q_{12}|}\sin \phi_q.
\label{eq:cpmix}
\end{eqnarray}
If $|q/p|_q=1$, $|B^{L,H}_q>$ would be also CP eigenstates.

Experimentally, $A^q_{CP}$ is obtained from the charge asymmetry in mixed
semileptonic $B^0_q$ decays:
\begin{eqnarray}
\nonumber
A^q_{CP}=A^q_{SL}=\frac{\Gamma(\bar{B^0_q} \rightarrow B^0_q\rightarrow \ell^+ \nu X)-\Gamma(B^0_q \rightarrow \bar{B^0_q}\rightarrow \ell^- \bar{\nu} X)}{\Gamma(\bar{B^0_q} \rightarrow B^0_q\rightarrow \ell^+ \nu X)+\Gamma(B^0_q \rightarrow \bar{B^0_q}\rightarrow \ell^- \bar{\nu} X)},
\end{eqnarray}
where $\ell$ means either electron or muon.

Standard Models predicts~\cite{nierste} 
$A^d_{SL}=(-4.0\pm 0.6)\times 10^{-4}$, $\phi_q=-4.9\degrees\pm 1.4\degrees$, 
$A^s_{SL}=(1.8\pm 0.3)\times 10^{-5}$ and  $\phi_s=0.24\degrees\pm 0.06\degrees$.
New particle exchange in the $B^0_q$ box diagrams could enhance $A^q_{SL}$ to 
values within the reach of the current precision of the experiments~\cite{lenz}.

Two classes of measurements are available: the inclusive dilepton asymmetry analyses and the flavor specific analyses.
In the first class, $A^q_{SL}$ is obtained from the dilepton asymmetry,
$A^b_{SL} = \frac{N(\ellp\ellp)-N(\ellm\ellm)}
{N(\ellp\ellp)+N(\ellm\ellm)} $, where an~\ellp
(\ellm) 
tags a \Bz (\Bzb).
Experiments at hadron colliders measure a combination of the 
$B^0_d$ and $B^0_s$ CP parameters, $A^b_{SL}=C_dA^d_{SL}+C_sA^s_{SL}$, where
the $C_{d,s}$ coefficients depend on the $B^0_{d,s}$ production rates and mixing
probabilities. Standard Model predicts $A^b_{SL}=(-2.8^{+0.5}_{-0.6})\times 10^{-4}$.
In the second class, $A^q_{SL}$ is obtained from the lepton charge asymmetry in the reconstructed 
$B^0_d \rightarrow D^{(*)} \ell X$ and $B^0_s \rightarrow D_s \ell X$ decays.

The current statistical precision of the experiments ($\order(10^{-3})$) requires a good control of the 
charge-asymmetric background originating from hadrons wrongly identified as leptons or
leptons from light hadron decays, and of the charge-dependent lepton identification
asymmetry that may produce a false signal. The systematic
uncertainties associated with the corrections for these effects 
constitute a severe limitation to the precision 
of the measurements.
These detector-related effects are reduced, when possible, by inverting the magnet polarities. 
They are estimated on control samples or determined simultaneously to $A^q_{SL}$.

\section{D0 like-sign dimuons charge asymmetry measurement}
\begin{figure}[htb]
\centering
\includegraphics[height=2.0in]{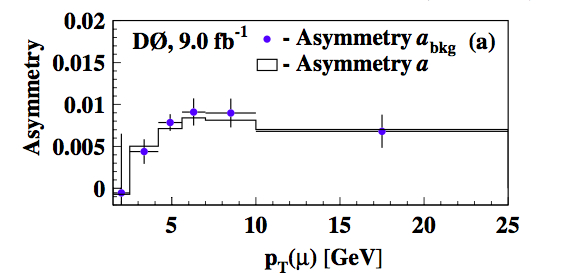}
\caption{Comparison between the expected background asymmetry (points with error bars) and the measured asymmetry
for the inclusive-muon sample (histogram) versus the muon transverse momentum.}
\label{fig:d01}
\end{figure}

From $9~fb^{-1}$ of $p\bar{p}$ collisions recorded at center-of-mass energy of $\sqrt{s}=1.96$~TeV~\cite{D0ll}, 
the semileptonic charge asymmetry $A^b_{SL}$ is measured from the inclusive
single muon and the like-sign dimuon charge asymmetries:
\begin{eqnarray}
\nonumber
 a_\mu=\frac{n(\mu^+)-n(\mu^-)}{n(\mu^+)+n(\mu^-)}, ~
A_{\mu \mu}=\frac{N(\mu^+ \mu^+)-N(\mu^- \mu^-)}{N(\mu^+ \mu^+)-N(\mu^- \mu^-)}.
\end{eqnarray}
Only about $3\%$ of the single muons and $30\%$ of the like-sign dimuons are produced 
in decays of mixed $B^0_q$ mesons.  
The contributions of muons from light hadrons decays, as well as their fractions and charge asymmetries,
are measured using $K^{*0}\rightarrow K^+ \pi^-$, $\phi\rightarrow K^+K^-$, $K_S\rightarrow \pi^+\pi^-$ and 
$\Lambda\rightarrow p\pi^-$ control samples. The contributions of muons from other sources is obtained from 
simulation. 
As shown in figure~\ref{fig:d01}, the observed single muon charge asymmetry agrees with the expectations
for the background. The asymmetry $A^b_{SL}$, extracted from the asymmetry of the inclusive-muon sample
taking into account the background contribution, is $A^b_{SL}=(-1.04\pm1.30 \stat\pm2.31 \syst)\%$, in agreement
with the Standard Model.
The systematic error is dominated by the uncertainty on the fraction of muons from kaon decays.

\begin{figure}[htb]
\centering
\includegraphics[height=2.0in]{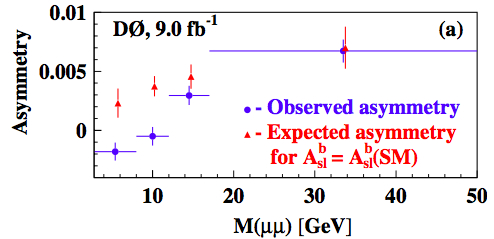}
\caption{The observed and expected like-sign dimuon charge asymmetries in bins of the dimuon invariant mass.}
\label{fig:d02}
\end{figure}

As shown in figure~\ref{fig:d02}, the observed like-sign dimuon asymmetry differs significantly from the 
background expectations. 
To reduce the dominant systematic uncertainty from the background fractions, which is correlated between the single muon and the dimuon
charge asymmetries, $A^b_{SL}$ 
is determined from a linear combination of $a_\mu$ and $A_{\mu\mu}$, 
$A^b_{SL}=(-0.787\pm0.172 \stat\pm0.093 \syst)\%$. This result differs by 3.9 standard deviations 
from the Standard Model prediction.

\begin{figure}[htb]
\centering
\includegraphics[height=2.0in]{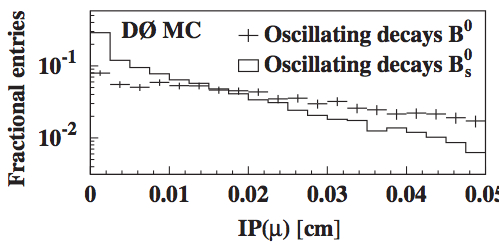}
\caption{The normalized impact parameter distribution for muons produced in decays of mixed $B^0_d$
and $B^0_s$ mesons.}
\label{fig:d03}
\end{figure}

The asymmetry $A^b_{SL}$ is produced by muons from direct semileptonic decays of $b$ quarks, 
characterized by a large impact parameter of their trajectories with respect to the primary vertex. 
The period of oscillation of the $B^0_d$ meson is many times longer than its lifetime so that the mixing
probability of $B^0_d$ increases with large impact parameters. The $B^0_s$ meson oscillates a number
of times within its lifetime so that it is fully mixed for any appreciable impact parameter
requirement. As a result, the fraction of mixed $B^0_d$ mesons can be increased by requiring a high
impact parameter, as shown in figure~\ref{fig:d03}. 
The measurements with large or small impact parameter use independent data samples,
and the dependence of $A^b_{SL}$ on $A^d_{SL}$ and $A^s_{SL}$ is different for the two samples.
The two $A^b_{SL}$ measurements in the regions with impact parameter larger or lower then $120~\mum$
can be therefore combined to obtain the values of $A^d_{SL}$ and $A^s_{SL}$: 
$A^d_{SL}=(-0.12\pm 0.52)\%$ and $A^s_{SL}=(-1.81\pm1.06)\%$.

\subsection{Interpretation}
The result for $A^b_{SL}$ is significantly different from the Standard Model expectation of CP violation
in mixing. The origin of this discrepancy is related to the dimuon like-sign charge asymmetry, whereas
the inclusive single muon charge asymmetry is in agreement with the expectations.
A search for any neglected sources of CP violation which could affect the dimuon like-sign
asymmetry leaving the single muon one uninfluenced is performed~\cite{D0int}.

The final states of the decays $B^0_d (\bar{B^0_d})\rightarrow c\bar{c} d\bar{d}$ are accessible from
both $B^0_d$ and $\bar{B^0}_d$. Therefore, the interference of decays to these final states
with and without $B^0$ mixing results in CP violation. It turns out that this CP violation reflects in 
a like-sign dimuon charge asymmetry, whereas the inclusive muon charge
asymmetry is not affected. The contribution of this process to the like-sign dimuon charge asymmetry 
is:
\begin{eqnarray}
\nonumber
 A(c\bar{c}d\bar{d})=-\sin(2\beta)\frac{x_d}{1+x^2_d}\omega(c\bar{c}d\bar{d})=-(0.045\pm0.016)\%,
\end{eqnarray}
where $x_d=\frac{\Delta m_d}{\Gamma_d}$, and $\omega(c\bar{c}d\bar{d})$ is the weight of this process
in the inclusive dimuon sample.

Taking into account this additional Standard Model source of dimuon charge asymmetry,
the D0 result becomes consistent with the Standard Model expectation within 3 standard
deviations. There is still however some room for new physics CP violation in $B^0_q$ mixing,
in the interference of $B^0_q$ decays with and without mixing, or in semileptonic decays
of $b$ and $c$ hadrons.

\section{Flavor Specific Analyses}
\subsection{D0 $A^q_{SL}$ measurement}
From $10.4~fb^{-1}$ of $p\bar{p}$ collisions recorded at center-of-mass energy of $\sqrt{s}=1.96$~TeV~\cite{D0fs1, D0fs2},
the flavor specific asymmetries $A^q_{SL}$ are measured using the exclusive decay channels 
$B^0_d\rightarrow D^-X\mu^+\nu~(D^-\rightarrow K^+\pi^-\pi^-)$, 
$B^0_d\rightarrow D^{*-}X\mu^+\nu~(D^{*-}\rightarrow \bar{D^0}\pi^-,~\bar{D^0}\rightarrow K^+\pi^-)$,
and $B^0_s\rightarrow D^-_sX\mu^+\nu~(D^-_s\rightarrow \phi \pi^-,~\phi\rightarrow K^+K^-)$.

With the assumption of no charge asymmetry in the $B^0_q$ meson production and no CP violation in charged $D$ mesons or in
$b$ semileptonic decays,
the CP violation asymmetries are obtained as:
\begin{eqnarray}
\label{eq:d0as}
A^q_{SL}=\frac{A^q-A^q_{BKG}}{F^{osc}_{B^0_q}},
\end{eqnarray}
where $A^q=\frac{N^q_{\mu^+ D^-}-N^q_{\mu^- D^+}}{N^q_{\mu^+ D^-}+N^q_{\mu^- D^+}}$ are the 
measured raw charge asymmetries, $A^q_{BKG}$ are
the detector-related asymmetries, and $F^{osc}_{B^0_q}$ are the fractions of signal events originating from oscillated
$B^0_q$, computed on simulation.

\begin{figure}[htb]
\centering
\includegraphics[height=2.0in]{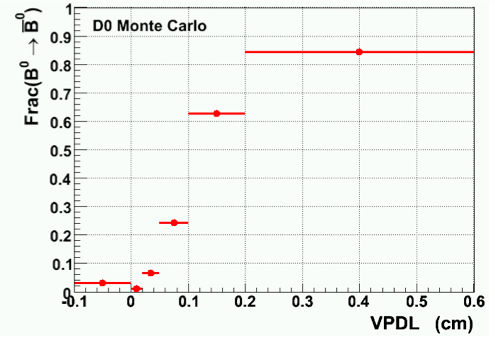}
\includegraphics[height=2.0in]{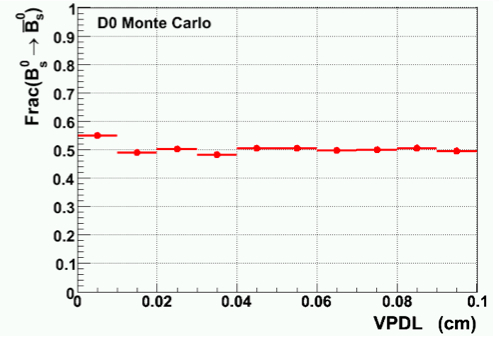}
\caption{Fraction of signal events from mixed $B^0_d$ mesons (left) and $B^0_s$ mesons (right) versus the visible proper decay length.}
\label{fig:d04}
\end{figure}

The visible proper decay length of the $B^0_q$ mesons is defined as $\mathrm{VPDL}=L_{xy}(B)\frac{cM(B)}{P_T(\mu D)}$, 
where $L_{xy}(B)$, $M(B)$ and $P_T(\mu D)$ are, respectively,
the transverse decay length, the reconstructed $B^0_q$ mass and the transverse momentum of the $\mu D$ system.
Due to the different oscillation periods of $B^0_d$ and $B^0_s$ mesons, the fractions $F^{osc}_{B^0_d}$ 
and $F^{osc}_{B^0_s}$ have a different
dependence on VPDL, as shown in figure~\ref{fig:d04}.
A time integrated analysis is used for the $A^s_{SL}$ measurement, whereas $A^d_{SL}$
is extracted by means of an analysis optimized in the different VPDL bins. 
The first two VPDL bins in the $B^0_d$ sample are populated only by background events, therefore they are used as a control sample. 

\begin{figure}[htb]
\centering
\includegraphics[height=1.9in]{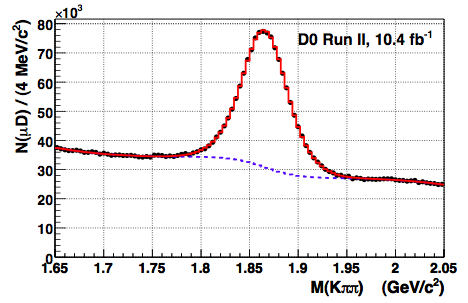}
\includegraphics[height=1.9in]{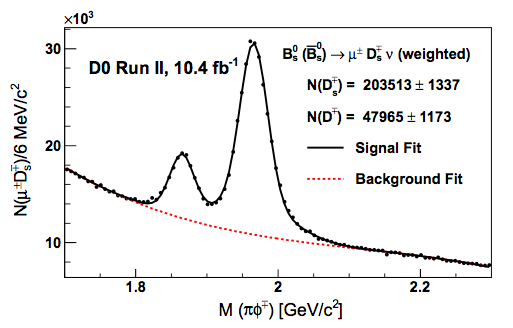}
\caption{Left: Invariant mass of $K\pi\pi$ candidates in $B^0_d$ decays.
Right: Invariant mass of $\pi\phi$ candidates in $B^0_s$ decays.}
\label{fig:d06}
\end{figure}

The $B^0_d$ and $B^0_s$ event selection is performed by means of two multivariate discriminant analyses exploiting
several kinematical and topological variables, as the reconstructed $D$ meson transverse decay length,
 the $B^0_q$ candidate mass and the track isolation. Final cuts are chosen to maximize the signal significance.
Figure~\ref{fig:d06} 
shows 
the $D$ and $D_s$ invariant mass for $B^0_d$ and $B^0_s$ decays.

The raw charge asymmetries $A^q$ are obtained from simultaneous fits to the sum and 
the difference of the $\mu^+ D^-$ and $\mu^- D^+$ invariant mass distributions.
A significant raw charge asymmetry $A^d=(1.48\pm0.41)\%$ is measured in the $B^0_d$ sample,
due to the different interaction lengths of the $K^+$ and $K^-$ mesons which result in a different reconstruction efficiency.

The corresponding asymmetry for the $B^0_s$ sample,
$A^s=(-0.40\pm0.33)\%$ is negligible due to the charge symmetric $K^+K^-$ final state.

The detector-related charge asymmetries from the muon identification efficiency and from kaon and pion 
decays depend on VPDL. They are estimated from $J/\psi\rightarrow \mu^+\mu^-$,
$K^{*0}\rightarrow K^+\pi^-$ and $K_S\rightarrow \pi^+\pi^-$ control samples 
and are reduced by reversing the magnet polarities every two weeks.

\begin{figure}[htb]
\centering
\includegraphics[height=2.0in]{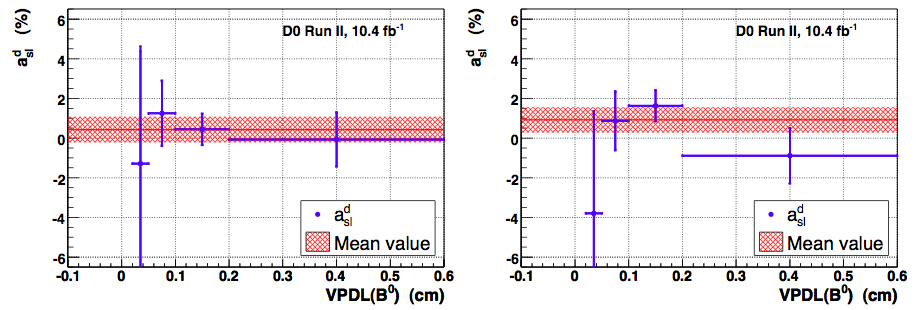}
\caption{$A^d_{SL}$ in bins of VPDL, for $\mu D$ (left) and $\mu D^*$ (right) samples. The cross-hatched bands show the
mean values and their total uncertainties.}
\label{fig:d08}
\end{figure}

Figure~\ref{fig:d08} shows the resulting $A^d_{SL}$ asymmetry versus VPDL for the $\mu D$ and $\mu D^*$ channels.
The final results from equation~\ref{eq:d0as} are $A^d_{SL}=(0.68\pm0.45\stat\pm0.14\syst)\%$ and 
$A^s_{SL}=(-1.12\pm0.74\stat\pm0.17\syst)\%$, in agreement with the Standard Model predictions. 
The systematics error is dominated by the uncertainties on the background
charge asymmetries and $F^{osc}_{B^0_q}$.

\subsection{LHCb $A^s_{SL}$ measurement}
From $1.0~fb^{-1}$ of $pp$ collisions recorded at center-of-mass energy of $\sqrt{s}=7$~TeV~\cite{LHCb},
the flavor specific asymmetry $A^s_{SL}$ is measured using the exclusive decay channel
$B^0_s\rightarrow D^-_sX\mu^+\nu~(D^-_s\rightarrow \phi \pi^-,~\phi\rightarrow K^+K^-)$.

$A^s_{SL}$ is obtained from the following equation:
\begin{eqnarray}
\label{eq:lhcbas}
A_{meas}=\frac{\Gamma[D^-_s\mu^+]-\Gamma[D^+_s\mu^-]}{\Gamma[D^-_s\mu^+]+\Gamma[D^+_s\mu^-]}
=\frac{A^s_{SL}}{2}+[A_p-\frac{A^s_{SL}}{2}]\frac{\int_0^\infty{e^{-\Gamma_s t}\cos(\Delta m_s t)\epsilon(t)dt}}
{\int_0^\infty{e^{-\Gamma_s t}\cosh\frac{\Delta m_s t}{2}\epsilon(t)dt}},
\end{eqnarray}
where the production asymmetry $A_p=\frac{N-\bar{N}}{N+\bar{N}}$,
defined in term of the number of produced particles
$N$ and antiparticles $\bar{N}$, is expected to be at most a few percent, and $\epsilon(t)$ is the decay 
time acceptance function for $B^0_s$ mesons.

Due to the rapid oscillations, the integral ratio is $0.2\%$  
and the effect of $A_p$ is reduced to the level of a few $10^{-4}$, under the goal of an error of the order
of $10^{-3}$.

The measured asymmetry is computed taking into account the detector effects, reduced by periodically
reversing the magnet polarities: 
\begin{eqnarray}
\nonumber
A_{meas}=\frac{N(D^-_s\mu^+)-N(D^+_s\mu^-)\times\frac{\epsilon(D^-_s\mu^+)}{\epsilon(D^+_s\mu^+)}}{N(D^-_s\mu^+)+N(D^+_s\mu^-)\times\frac{\epsilon(D^-_s\mu^+)}{\epsilon(D^+_s\mu^+)}}.
\end{eqnarray}
The relative efficiencies $\frac{\epsilon(D^-_s\mu^+)}{\epsilon(D^+_s\mu^+)}$ are computed on calibration samples:
the track efficiency ratio $\epsilon(\pi^+)/\epsilon(\pi^-)$ is obtained from the ratio of fully reconstructed and 
partially reconstructed $D^{*+}\rightarrow D^0\pi^+, D^0\rightarrow K^-\pi^+\pi^-(\pi^+)$,
and the muon efficiency ratio $\epsilon(\mu^+)/\epsilon(\mu^-)$ from a sample of reconstructed
$J/\psi\rightarrow \mu^+\mu^-$ decays, using a tag and probe method.

\begin{figure}[htb]
\centering
\includegraphics[height=1.6in]{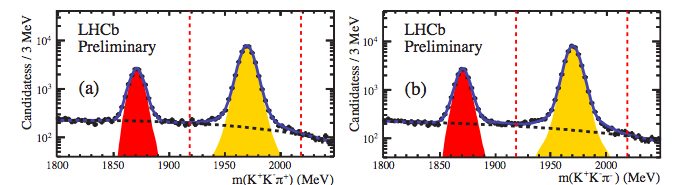}
\caption{Invariant mass distributions for (a) $K^+K^-\pi^+$ candidates and (b) $K^+K^-\pi^-$ for magnet up
with $m(KK)$ within $\pm 20$~MeV of the $\phi$ meson mass.}
\label{fig:lhcb1}
\end{figure}

Figure~\ref{fig:lhcb1} shows the selected $D_s$ signal yields for the two different charge combinations.
The background asymmetry due to kaon and pion misidentification, prompt charm decays, $B\rightarrow D_s X$
and $B\rightarrow D_s K\mu\nu X$ decays is estimated to be of $\order(10^{-4})$. 

\begin{figure}[htb]
\centering
\includegraphics[height=1.8in]{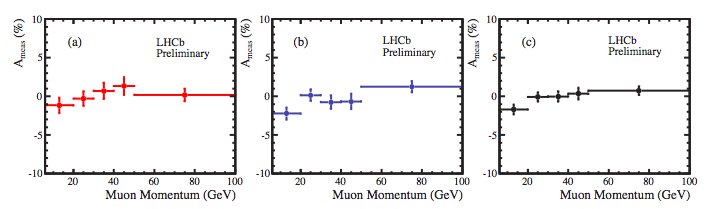}
\caption{Corrected $A_{meas}$ as a function of the muon momentum for (a) magnet up, (b) magnet down, and (c) the average.}
\label{fig:lhcb2}
\end{figure}

Figure~\ref{fig:lhcb2} shows the corrected measured asymmetry as a function of the muon momentum.
The final results from equation~\ref{eq:lhcbas} is
$A^s_{SL}=(-0.24\pm0.54\stat\pm0.33\syst)\%$, in agreement with the Standard Model predictions. 
The systematics uncertainty is dominated by the statistical error on the muon efficiency ratio 
$\epsilon(\mu^+)/\epsilon(\mu^-)$.

\subsection{Babar $A^d_{SL}$ measurement}
From $425.7 fb^{-1}$  of $e^+ e^-$ collisions recorded at the $\Upsilon(4S)$~\cite{babar},
the flavor specific asymmetry $A^d_{SL}$ is measured using  
the semileptonic transition
$B^0_d\rightarrow D^{*-} X\ell^+ \nu$, with a partial reconstruction of the $(D^{*-}\rightarrow \pi^- \bar{D^0} )$ decay.

The observed asymmetry between the number of events with an $\ell^+$ compared to those with an $\ell^-$
is: 
\begin{eqnarray}
A_\ell\simeq A_{r\ell}+A^d_{SL}\chi_d,
\label{eq:all}
\end{eqnarray}
where $\chi_d$ is the integrated mixing probability
for $B^0_d$ mesons, and $A_{r\ell}$ is the detector-induced charge asymmetry in the $B^0_d$ reconstruction.

The flavor of the other $B^0_d$ is tagged looking for charged kaons in the event $(K_T)$. A state decaying as
a $B^0_d~(\bar{B^0_d})$ meson results most often in a $K^+~(K^-)$. The observed asymmetry
in the rate of mixed events is:
\begin{eqnarray}
A_T=\frac{N(\ell^+ K^+_T)-N(\ell^- K^-_T)}{N(\ell^+ K^+_T)-N(\ell^- K^-_T)}\simeq A_{r\ell}+A_K+A^d_{SL},
\label{eq:btag}
\end{eqnarray}
where $A_K$ is the detector charge asymmetry in kaon reconstruction.
A kaon with the same charge as the $\ell$ might also come from the Cabibbo-Favored decays of the $D^0$
meson produced with the lepton from the partially reconstructed side $(K_R)$. The asymmetry observed for these events is:
\begin{eqnarray}
A_T=\frac{N(\ell^+ K^+_R)-N(\ell^- K^-_R)}{N(\ell^+ K^+_R)-N(\ell^- K^-_R)}\simeq A_{r\ell}+A_K+A^d_{SL}\chi_d.
\label{eq:dtag}
\end{eqnarray}
Eqs.~\ref{eq:all}, \ref{eq:btag} and \ref{eq:dtag} can be used to extract $A^d_{SL}$ and the detector induced
asymmetries. Due to the small lifetime of the $D^0$ meson, the separation in space between the
$K_R$ and the $\ell \pi$ production points is much smaller than for $K_T$. Therefore the proper time difference $\Delta t$
between the two $B^0_d$ meson decays is used as a discriminant variable.
Kaons in the $K_R$ sample are usually emitted in the hemisphere opposite to the
$\ell$, while $K_T$ are produced randomly, so in addition the cosine of the
angle $\theta_{\ell K}$ between the lepton and the kaon is used.

\begin{figure}[htb]
\centering
\includegraphics[height=2.5in]{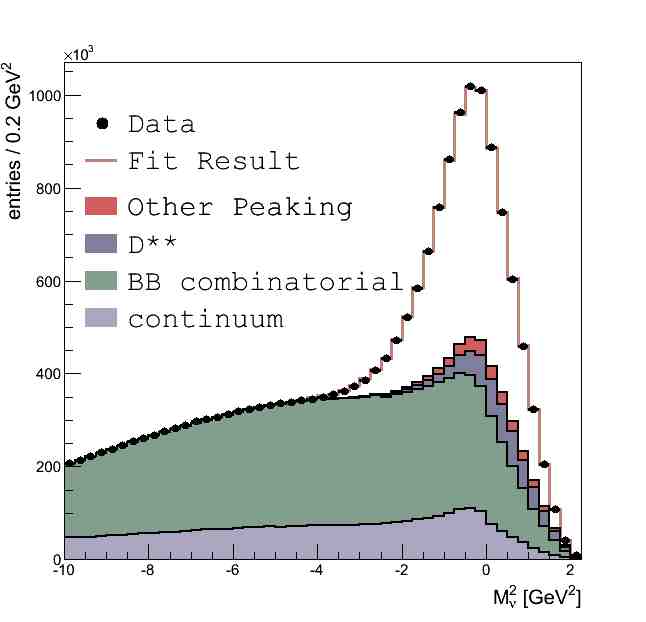}
\caption{$M^2_\nu$ distribution for selected events. The data are represented by the points with
error bars. The fitted contributions are overlaid.}
\label{fig:babar1}
\end{figure}

The $B^0_d\rightarrow D^{*-} X\ell^+ \nu~(D^{*-}\rightarrow \pi^- \bar{D^0} )$ events are selected searching for combinations of a charged
lepton and a low momentum pion with opposite charge, consistent with originating from a common
vertex. Signal is selected by means of the squared of the unobserved neutrino mass, $M^2_\nu=(E_{beam}-E_{D^*}-E_\ell)^2-(\vec{p_{D^*}}+\vec{p_\ell})^2$ where the $B^0_d$ energy is identified with the beam energy
$E_{beam}$ in the $e^+ e^-$ center of mass frame, $E_\ell$ and $\vec{p_\ell}$ are energy and momentum
vector of the lepton, and $p_{D^*}$ is the estimated $D^*$ momentum vector. The signal fraction 
is determined by fitting the $M^2_\nu$ distribution with the sum of continuum, combinatorial and peaking
events. The shape for combinatorial and peaking events is estimated from the simulation, whereas the shape for continuum is fixed to the shape 
of events collected 40~MeV below the $\Upsilon(4S)$ resonance. Figure~\ref{fig:babar1} shows the result of the $M^2_\nu$ fit.

\begin{figure}[htb]
\centering
\includegraphics[height=2.5in]{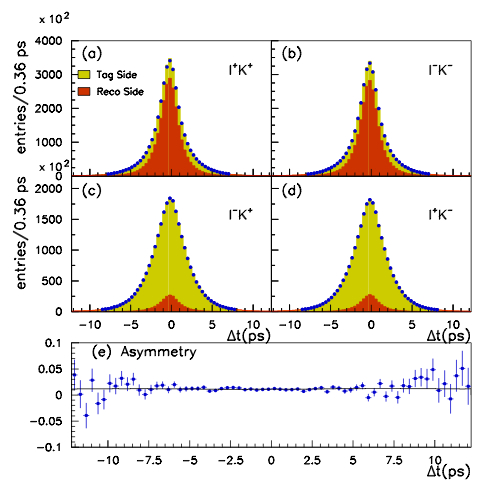}
\includegraphics[height=2.5in]{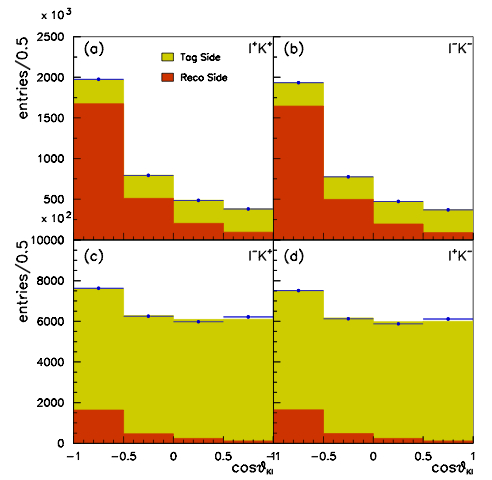}
\caption{Distribution of $\Delta t$ (left) and of $\cos(\theta_{\ell K})$ (right) for the continuum subtracted data (points with error bar)
and fitted contributions from $K_R$ (dark) and $K_T$ (light) for all the $K\ell$ charge combinations.
The raw asymmetry between $K^+\ell^+$ and $K^-\ell^-$ is shown in the left plot.}
\label{fig:babar2}
\end{figure}

$A^d_{SL}$ is obtained with a binned fit to $\Delta t$ and $\cos(\theta_{\ell K})$. The result of the fit is shown in
figure~\ref{fig:babar2}.
The final result is $A^d_{SL}=(0.06\pm0.17^{+0.38}_{-0.32})\%$, in agreement with the Standard Model predictions.
The systematic error is dominated by the uncertainty on the sample composition.

\section{Conclusions}
\subsection{World Averages}
\begin{figure}[htb]
\centering
\includegraphics[height=2.5in]{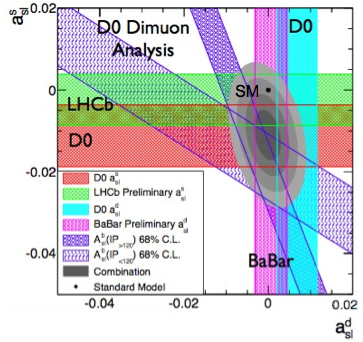}
\includegraphics[height=2.5in]{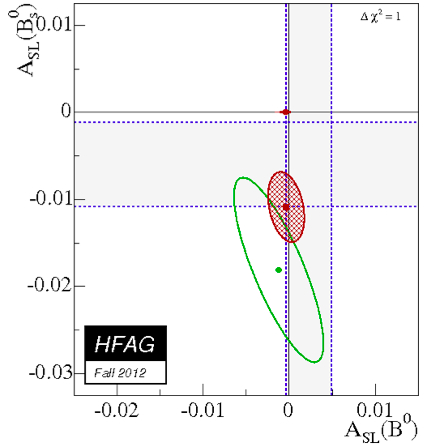}
\caption{Results of the various $A^q_{SL}$ measurements in the $(A^s_{SL},A^d_{SL})$ plane (left). 
World average from the 2-dimensional fit computed by the HFAG 
compared with the Standard Model predictions (right). The vertical and horizontal bands are the averages of the result of the flavor specific
$A^d_{SL}$ and $A^s_{SL}$ measurements, respectively.
The green ellipse is the D0 measurement with same-sign
dimuons and the red ellipse is the result of the 2-dimensional average. The red point close to (0,0) is the Standard
Model prediction.}
\label{fig:con1}
\end{figure}

Figure~\ref{fig:con1} shows the results of the analyses described in the previous sections~\cite{bertram}, and the world average
computed by the Heavy Flavor Averaging Group
from a 2-dimensional fit in the plane $(A^s_{SL}, A^d_{SL})$, compared with the Standard Model predictions~\cite{hfag}. 

The average result from the 
B-factories measurements for the $B^0_d$ meson is:
$A^d_{SL}=(0.02\pm0.31)\%$. 
The world averages of the flavor specific analyses for the $B^0_d$ and $B^0_s$ mesons are:
$A^d_{SL}=(0.23\pm0.26)\%$ and $A^s_{SL}=(-0.60\pm0.49)\%$, respectively, in agreement with the Standard Model
predictions. The result of the 2-dimensional fit is 
\begin{eqnarray}
\nonumber
A^d_{SL}=(-0.03\pm0.21)\%\\
\nonumber
A^s_{SL}=(-1.09\pm0.40)\%
\end{eqnarray}
which, due to the D0 dilepton measurement, deviates by 2.4 standard deviations from the expectations.

\subsection{Constraints on New Physics}
The relations between the off-diagonal matrix elements $|M^q_{12}|$ and
$|\Gamma^q_{12}|$, the CP violating phase $\phi_q$ and the observables $\Delta m_q$, 
$\Delta\Gamma_q$ and $A^q_{SL}$ have been already discussed in section~\ref{sec:intro}.
$M^q_{12}$ is very sensitive to new physics contributions~\cite{leni}. Therefore, 
the two complex parameters $\Delta_s$ and $\Delta_d$, defined as $M^{NP,q}_{12}=M^{SM,q}_{12}\Delta_q$,
$\Delta_q=|\Delta_q|e^{i\phi^\Delta_q}$ can differ from the Standard Model value
$\Delta_q=1$, resulting in a modified semileptonic asymmetry 
$A^{NP,q}_{SL}=\frac{|\Gamma^q_{12}|}{|M^{SM,q}|}\frac{\sin(\phi^{SM}_q+\phi^\Delta_q)}{|\Delta_q|}$.

The new phases $\phi^\Delta_q$ also shift the CP phases extracted from the mixing-induced
CP asymmetries in $B^0_d\rightarrow J/\psi K$ and $B^0_s\rightarrow J/\psi \phi$ from 
$2\beta$ to $2\beta+\phi^\Delta_d$ and from $2\beta_s$ to $2\beta_s-\phi^\Delta_s$,
respectively. Therefore the recent LHCb results on $B^0_s\rightarrow  J/\psi \phi$~\cite{lhcb2}
can be used to constraint the new physics phase $\phi^\Delta_s$.

\begin{figure}[htb]
\centering
\includegraphics[height=2.5in]{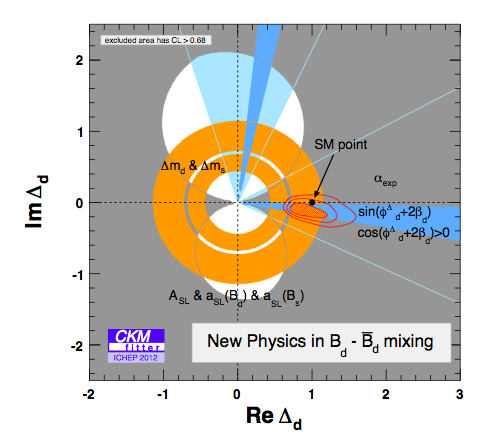}
\includegraphics[height=2.5in]{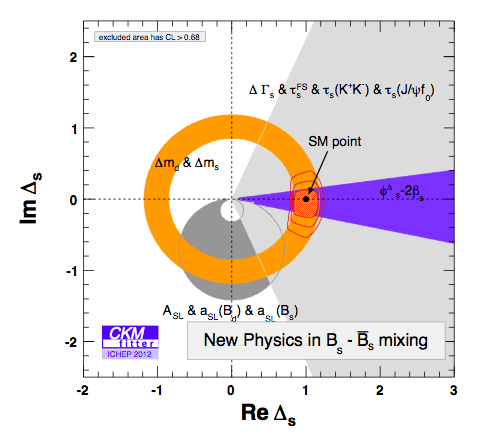}
\caption{Allowed regions for the new physics parameters $\Delta_d$ (left) and $\Delta_s$ (right).}
\label{fig:con2}
\end{figure}

Figure~\ref{fig:con2} shows the result of the global fits of $\Delta_q$ to all the relevant data in the plane 
$(\Im \Delta_q, \Re \Delta_q)$~\cite{nierste, ckmf}.
Due to the LHCb constraint on $\phi^\Delta_s$, the Standard Model predictions are currently disfavored by only 1
standard deviation and the new physics scenario described above cannot
explain the dimuon D0 result anymore. 

\subsection{Summary}
CP violation in the mixing of $B^0_q$ mesons is an excellent laboratory for the search
for physics beyond the Standard Model. In the last couple of years five new measurements
have been performed by B-Factories and Hadron Collider experiments, with an experimental
precision of the order of $10^{-3}$. All the results, but the D0 measurement with dimuons, 
are in agreement with the Standard Model expectations.

In the near future, the study of CP violation in the $B^0_q$ mixing at the LHC and at
the high intensity B-factories will offer the opportunity to improve the experimental
techniques, perform very stringent Standard Model test and, hopefully, 
to discover or to understand new physics.






\begin{thebibliography}{99}



\bibitem{nierste}
U. Nierste, arXiv:1212.5805 (2012).

\bibitem{lenz}
A. Lenz, U. Nierste, Journ. High En. Phys.~{\bf 0706}, 072 (2007).

\bibitem{D0ll}
The D0 Collaboration, Phys. Rev. D~{\bf 84}, 052007 (2011).

\bibitem{D0int}
G. Borissov and B. Hoeneisen, Phys. Rev. D~{\bf 87}, 074020 (2013).

\bibitem{D0fs1}
The D0 Collaboration, Phys. Rev. D~{\bf 86}, 072009 (2012).

\bibitem{D0fs2}
The D0 Collaboration, Phys. Rev. Lett.~{\bf 110}, 011801 (2013).

\bibitem{LHCb}
The LHCb Collaboration, LHCb-CONF-2012-022 (2012).

\bibitem{babar}
The Babar Collaboration, arXiv:1305.1575 (2013), submitted to Phys. Rev. Lett.

\bibitem{bertram}
I. Bertram, "Recent B-Physics Results from the D0 Experiment", presented at the 
Deep Inelastic Scattering Conference, DIS 2013.

\bibitem{hfag}
The Heavy Flavor Averaging Group, http://www.slac.stanford.edu/xorg/hfag/

\bibitem{leni}
A. Lenz et al., Phys. Rev. D~{\bf 86}, 033008 (2012).

\bibitem{lhcb2}
The LHCb Collaboration, LHCb-CONF-2012-002.

\bibitem{ckmf}
The CKMfitter Group, http://ckmfitter.in2p3.fr

\end{thebibliography}
\end{document}